 \documentclass[prl,aps,showpacs,tightenlines,twocolumn]{revtex4}

 \usepackage{graphicx}
  
\newcommand{\bear}{\begin{array}}  \newcommand{\eear}{\end{array}}
\newcommand{\bea}{\begin{eqnarray}}  \newcommand{\eea}{\end{eqnarray}}
\newcommand{\beq}{\begin{equation}}  \newcommand{\eeq}{\end{equation}}
\newcommand{\bef}{\begin{figure}}  \newcommand{\eef}{\end{figure}}
\newcommand{\bec}{\begin{center}}  \newcommand{\eec}{\end{center}}
\newcommand{\non}{\nonumber}

\newcommand{\la}{\left\langle} \newcommand{\ra}{\right\rangle}

\newcommand{\gtrsim}{ \mathop{}_{\textstyle \sim}^{\textstyle >} }
\newcommand{\lesssim}{ \mathop{}_{\textstyle \sim}^{\textstyle <} }

\def\EQ#1{Eq.~(\ref{#1})}
\def\EQS#1{Eqs.~(\ref{#1})}

\begin{document}

\title{Late-time entropy production due to the decay of domain walls}
\author{Masahiro Kawasaki}

\author{Fuminobu Takahashi}
\affiliation{Institute for Cosmic Ray Research, 
University of Tokyo, Kashiwa 277-8582, Japan}

\date{\today}
\begin{abstract}
  It is shown that late-time decay of domain walls can
dilute unwanted relics such as moduli, if the universe was
dominated by frustrated domain walls with tension
 $\sigma = (1\sim 100 \,{\rm TeV})^3$.
 Since energy density of the frustrated domain walls  decreases
as slow as the inverse of the scale factor,
an overclosure limit on the axion decay constant $f_a$
is also considerably relaxed. In fact $f_a$ can be as large
as the Planck scale, which may enable us to naturally implement the
QCD axion in the string scheme. Furthermore, in contrast to
thermal inflation models, the Affleck-Dine baryogenesis can generate
enough asymmetry to explain the present baryon abundance, even
in the presence of late-time entropy production.
\end{abstract}

\maketitle

Particle physics beyond the standard model predicts a number of new
particles, some of which have long lifetimes and decay at cosmological
time scales. If such long-lived particles are substantially produced 
in the early universe, they may give crucial effects on cosmology and
spoil success of the standard big-bang model. 

One well-known example of such dangerous relic particles is gravitino 
in supergravity theories. Gravitinos are produced during reheating 
after inflation and the abundance of the produced gravitino is 
proportional to the reheating temperature. 
The gravitino with mass $\sim 0.1-10$~TeV decays 
during or after the big-bang nucleosynthesis (BBN) and may 
destroy the  light elements synthesized in BBN. 
Thus, in order to keep success of BBN the reheating temperature 
should be sufficiently low (for recent works, see 
\cite{Kawasaki:2000qr,Jedamzik:1999di,Cyburt:2002uv,
Kawasaki:2004qu}). 

Another class of dangerous relics are light scalar fields called 
moduli which appear in superstring theories~\cite{Coughlan:1983ci,
Banks:1993en,deCarlos:1993jw}. The moduli fields are expected to 
acquire masses of the order of the gravitino mass from 
nonperturbative effects of the SUSY breaking and have long lifetimes 
since they have only the gravitationally suppressed interactions.
During inflation the field value of moduli generally deviates from
the vacuum value due to extra SUSY breaking by the cosmic density.
The deviation is of the order of the Planck scale $M_{G} (=2.4\times 10^{18}~{\rm GeV})$ since 
no other mass scale exists. Then, when the Hubble becomes 
less than the modulus mass, the modulus field starts to osciilate 
with amplitude $\sim M_{G}$. Because the modulus denity is comparable
to the cosmic density at onset of the oscillation, it soon dominates
the density of the universe and causes the serious cosmological 
problem (moduli problem). Since the abundance of the moduli
is independent of the reheating temperature, the moduli problem
is much more serious than the gravitino problem. 
To solve the moduli problem (as well as the gravitino problem) 
it is necessary to dilute the oscillating moduli  by huge entropy 
production.     
So far the most successful model to produce such large entropy
is thermal inflation proposed by Lyth and Stewart~\cite{Lyth:1995hj,
Lyth:1995ka}. The thermal inflation is mini-inflation with the number of
e-folds $\sim 10$ which takes place at the weak scale. A comprehensive
study of the thermal inflation was done in Ref.~\cite{Asaka:1999xd}
and it was shown that for modulus mass between $10$~eV and $10$~TeV
the thermal inflation solves the cosmological moduli problem. 
However, the thermal inflation also dilute the baryon number of 
the universe and hence we need an efficient baryogenesis mechanism.
In Refs.~\cite{deGouvea:1997tn,Asaka:1999xd} it was pointed out 
that the Affleck-Dine mechanism does that job. 
But later it was shown that $Q$-ball formation
obstructs the baryogenesis~\cite{Kasuya:2001tp}. 
 
In this letter we propose an alternative model to dilute the 
dangerous cosmological relics. In our model domain walls, which are
usually considered as cosmological disaster, play an important role.
The domain walls are produced when some discrete symmetry is 
spontaneously broken. The domain wall network in the universe
becomes very complicated if  there are many discrete vacua.
The density of such domain wall network decreases very slowly
and quickly dominates the universe. In extreme case, the domain walls
become frustrated and their density $\rho_{DW}$ decreases 
as $\rho_{DW}\propto a^{-1}$ ($a$: the scale factor) which is 
slower than the scaling evolution $\rho_{DW}\propto a^{-3/2}$ 
in matter-dominated universe. Furthermore, if the scalar 
potential has a tiny term which explicitly breaks the discrete 
symmetry, the vacua separated by domain walls have slightly different 
energies and hence the domain wall network is no longer stable and 
decays producing large entropy. It will be shown that 
the entropy production by the decay of domain wall is enough to 
dilute the moduli density. Moreover, because the density of
domain wall decreases much more slowly than the matter density,
the cosmic axion density is also diluted. As a result, the axion
decay constant $f_{a}$ as large as $M_{G}$ is allowed, which might
enable us to implement the axion into the string framework. 
In addition, the Affleck-Dine baryogenesis is compatible with our
model as we will see below. 
Therefore, the domain walls provide more efficient and consistent 
dilution mechanism than the thermal inflation.

First we present our scenario that  the decay of
domain walls generates large entropy to dilute unwanted
relics such as moduli and gravitinos, and derive constraints
on  tension of  the domain walls. In contrast to the other
candidates for late-time entropy production,  domain walls
have an advantage that the energy density decreases relatively
slowly, which ensures huge entropy production. To be specific,
an equation-of-state $w$ equals to $-2/3$ if the domain walls
are completely frustrated, that is, the structure of the domain wall 
network remains unchanged as the universe expands.
Throughout this letter, we assume this is the case. Later
we give a model which would lead to such frustrated
domain walls.

Let us consider the evolution of the energy density of  domain walls, 
which eventually dominate  the universe. 
The energy density of the domain walls at the formation 
is estimated as
\beq
\rho_{DW,i} \sim \sigma H_i,
\eeq
where $\sigma$ is the tension of the wall, $H_i$ the Hubble parameter
at the phase transition. We have assumed that there are only a few
domain walls per one horizon when they are formed. Here and in what
follows we neglect $O(1)$ numerical factors. If the domain
walls are fully frustrated, the energy density falls off as the
inverse of the scale factor: $\rho_{DW} \propto a^{-1}$. 
For  definite discussion, we assume that
the universe is dominated by either the inflaton or modulus fields when
the domain walls are formed. Noting that the energy density of an 
oscillating massive scalar field decreases as $ \propto a^{-3}$, 
the domain walls 
dominate the universe when the Hubble parameter becomes equal
to
\beq
\label{eq:heq}
H_{eq} \sim \sigma^{\frac{3}{4}} H_i^{\frac{1}{4}} M_G^{-\frac{3}{2}}.
\eeq
If the reheating occurs earlier than $H=H_{eq}$, the radiation-dominated epoch 
might last for a while, which makes it easy for the domain walls to
dominate the unverse. Thus, for conservative discussion, we assume 
that the universe is dominated by nonrelativistic particles until 
the domain walls  dominate the universe.

The decay of domain walls can proceed
if there is a tiny bias between vacua, $\delta \rho$. When this energy difference
becomes comparable to the energy of the domain walls, the decay occurs and  the universe is reheated. Since the universe was dominated by the domain walls
in our scenario, $\delta \rho$ is simply related to the decay temperature as
$\delta \rho \sim T_d^4$.
In order to have both large enough entropy production and the successful
BBN, we take the decay temperature as low as $10{\rm MeV}$.
Note that this value is just an exemplified value and that the successful
dilution is actually realized for a wide range of the decay temperature,
say, $T_d = 10 {\rm MeV} \sim 10{\rm GeV}$ (see \EQS{eq:sigma_u},
(\ref{eq:sigma_d1}) and (\ref{eq:sigma_d2})).
It is the smallness of $\delta \rho$ that enables domain walls to
be long-lived. Let us note that, when the domain walls decay, 
at least several domain walls must be present inside one horizon, otherwise the old inflation occurs somewhere leading to unacceptably inhomogeneous universe.
This requires  that the decay temperature $T_d$  satisfy the following inequality:
$ T_d^4  > \sigma H_d$,
where $H_d \sim T_d^2/M_G$ is
the Hubble parameter when the domain walls decay.
Thus the tension is bounded above:
\beq
\label{eq:sigma_u}
\sigma <T_d^2 M_G \sim 
\left(100 {\rm \,TeV}\right)^3 \left(\frac{T_d}{10{\rm \, MeV}}\right)^{2}.
\eeq
On the other hand, the tension should be large enough to dominate the
universe before the decay: $H_{eq} > H_d$. That is, 
\beq
\label{eq:sigma_d1}
\sigma >  \left(\frac{T_d^8 M_G^2}{H_i} \right)^{\frac{1}{3}} 
            \sim  \left(100 {\rm \,GeV}\right)^3 \left(\frac{T_d}{10{\rm \, MeV}}\right)^{\frac{8}{3}} \left(\frac{H_i}{1{\rm \, TeV}}\right)^{-\frac{1}{3}}.
\eeq

Now let us estimate the the abundance of the modulus field after decay of
the domain walls. For domain walls with the tension satisfying
(\ref{eq:sigma_u}) and (\ref{eq:sigma_d1}),  $H_{eq}$ is much smaller
than the modulus mass. 
Since the modulus field dominates
 the universe as soon as it starts oscillating,  it is reasonable to assume that 
the universe was dominated by the modulus field when $H=H_{eq}$. 
Then the modulus-to-entropy ratio is
given by
\bea
\frac{\rho_{mod}}{s} 
	\sim T_d   \left(\frac{H_d}{H_{eq}}\right)^4
		                \sim \frac{T_d^9 M_G^2}{\sigma^3 H_i},
\eea
where we have used the fact that the Hubble parameter falls off as
$\propto a^{-1/2}$ while the domain walls dominate the universe,
and Eq.~(\ref{eq:heq}) is substituted in the
last equation.
This must satisfy the observational bound:
\beq
\frac{\rho_{mod}}{s} < \kappa \, \frac{\rho_c}{s} = 
		3.6 \times 10^{-9} \kappa \, h^2   {\rm \,GeV},
\eeq
where $\rho_c$ is the critical density, $\kappa$ varies from
$10^{-10}$ to $0.2$ depending on the modulus 
mass~\cite{Asaka:1999xd}, and 
$h$ is  the Hubble constant in units of 100\ km/sec/Mpc.
Thus the tension is further constrained as
\beq
\label{eq:sigma_d2}
\sigma > \kappa^{-\frac{1}{3}} (500 {\rm\, GeV})^3 
		\left(\frac{T_d}{10{\rm \, MeV}}\right)^3 
		 \left(\frac{H_i}{1{\rm \,TeV}}\right)^{-\frac{1}{3}}. 
\eeq
To sum up, the domain walls with the tension $\sigma = (1 \sim 100 {\rm\, TeV})^3$,
which decay just before the BBN starts, can dilute the moduli to the
observationally allowed level.

Next we present a model that would lead to
formation of the frustrated domain walls. Let us consider
the following superpotential:
\beq
W = \sqrt{\lambda} \sum_{i}^N \sum_j^N Z_{ij} \Phi_i \Phi_j + \frac{2\epsilon}{\sqrt{\lambda}} \sum_i Z_{ii} \Phi_i^2,
\eeq
where $Z_{ij}$ and $\Phi_i$ are gauge-singlet superfields, and real coupling
constants $\lambda >0$ and $\epsilon$ are assumed to 
satisfy the inequality, $\lambda \gg |\epsilon|$~\footnote{
Note that large hierarchy between $\lambda$ and $\epsilon$ is not necessary,
since the tension of the domain walls is independent of $\epsilon$ as long as
$\epsilon$ is not too small. In fact, $\epsilon \sim 0.1 \lambda$ does work.
}.
These interactions respect $S_N$ symmetry if $\Phi_i$ and $Z_{ij}$ are taken to 
be the fundamental and bi-fundamental representations of $S_N$ permutation group.
 The $R$-charges are assigned as $R_\Phi = 0$ 
and $R_Z = 2$. We assume that $Z_{ij}$ always sits at the origin: $\la Z_{ij} \ra=0$,
which can be realized if $Z_{ij}$ acquires a positive Hubble-induced mass
term during relevant epoch.

Then we obtain the following effective potential for $\Phi_i$, 
\bea
\label{eq:pote1}
V(\Phi)& \simeq& V_0 - m_0^2 \sum_i |\Phi_i|^2- m_{3/2}^2 \sum_i
(\Phi_i^2 + \Phi_i^*{}^2) \non\\
&&+ \lambda (\sum_i |\Phi_i|^2)^2+
4\epsilon \sum_i|\Phi_i|^4,
\eea
where $m_{3/2}$ is the gravitino mass, $V_0 = O(m_0^4/\lambda)$
is chosen in such a way that the cosmological constant vanishes in the true
minimum, and we have used $\lambda \gg | \epsilon|$.  Here we have
assumed the negative mass of the order of the weak scale 
at the origin, $m_0\sim \Lambda_{EW} \sim 1{\rm \,TeV}$,
which is induced by the SUSY breaking effects.
The third term comes from the gravity-mediated SUSY
breaking effects. Due to this term, the global minima of the
potential lie along the real axes of $\Phi_i$. 
In order to study the vacuum structure
of this potential, let us concentrate on the
the real components of $\Phi_i$ by setting ${\rm Im}\,\Phi_i =0$.
Rewriting the potential in terms of the real components $\phi_i \equiv
\sqrt{2}\,{\rm Re} \,\Phi_i$, we obtain
\beq
\label{eq:pote2}
V(\phi) \simeq V_0 - m_0^2 \sum_i \phi_i^2/2
+ \lambda (\sum_i \phi_i^2)^2 /4+
\epsilon \sum_i \phi_i^4,
\eeq
where the third term in \EQ{eq:pote1} is neglected since we are interested
in the case of $m_{3/2} \lesssim  m_0$. 
The scalar potential given by \EQ{eq:pote2} actually agrees with the $O(N)$ 
model studied in Refs.~\cite{Kubotani:1991kw,Ishihara:1991cd}. The approximate
$O(N)$ symmetry originates from the hierarchical couplings, $\lambda$ and $\epsilon$.
Note that this potential is obtained by disregarding ${\rm Im} \,\Phi_i$, so it does not necessarily  
give a right answer, for instance,
when one estimates the energy density inside domain walls,
although it is still useful to study the vacuum structure.

The positions of the global minima of $V(\phi)$ depend on
the sign of $\epsilon$. For $\epsilon <0$, there are $2N$ minima:
\beq
\phi_{{\rm min}}^{\pm \,(i)} = (0,\dots,0,\pm v_1,0,\dots,0),
~~~~~{\rm for}~~~i=1\sim N
\eeq
with $v_1 \equiv m_0/\sqrt{\lambda}$.
On the other hand, if $\epsilon > 0$, the potential minima are given by
\beq
\label{eq:vac2}
\phi_{{\rm min}} = ( \pm v_2,\dots,\pm v_2)
\eeq
with $v_2 \equiv m_0/\sqrt{\lambda N}$,
where  arbitrary combination of the signs are allowed. In the following, 
we consider the case of $\epsilon > 0$. Then
all $Z_{ij}$ fields acquire masses of the order of $v_2 \sim \Lambda_{EW}$  in these minima. Also,
the number of the minima  is $2^N$.
Thus, for large $N$,  the vacuum structure becomes more complicated, compared to the former case. Since there are many types of vacua, 
pair annihilation  processes of walls are expected to be highly 
suppressed, which enables
walls to deviate from the scaling law. The minima are separated by the potential barrier represented
by the third term in Eq.~(\ref{eq:pote1}), 
if $\epsilon$ is larger than $m_{3/2}^2/v_2^2 \sim \lambda 
N m_{3/2}^2 /m_0^2$. Then
the tension of the walls is given by
\beq
\label{eq:tension1}
\sigma_{1} \sim m_{3/2} v_2^2 \sim ( 1{\rm \, TeV} )^3\left(\lambda N \right)^{-1}
 \left(\frac{m_{3/2}}{1 {\rm \, TeV}}\right)
					 \left(\frac{m_{0}}{1 {\rm \, TeV}}\right)^2.
\eeq
 On the other hand, if 
$\epsilon$ is smaller than $m_{3/2}^2/v_2^2$, the potential barrier is the
last term in Eq.~(\ref{eq:pote1}). The tension is
$\sigma_{2} \sim 
|\epsilon|^{\frac{1}{2}} v_2^3  \leq \sigma_1,$
where the last equality holds when $\epsilon \sim m_{3/2}^2/v_2^2$.
Thus the tension of the domain walls in this model can satisfy the
bound derived in the previous section, although  $\lambda N \lesssim 1$ 
might be necessary in the latter case.

We assume that all $\Phi_i$ get  positive Hubble-induced masses
therefore sit at the origin until the Hubble parameter becomes comparable
to $m_0$. Then those scalar fields start rolling down and get settled 
somwhere on $S^{N-1}$ defined by $\sum_i |\Phi_i|^2 = 2 N v_2^2$.
When the Hubble parameter becomes equal to the gravitino mass,
all $\Phi_i$ move to the real axes due to the third term in \EQ{eq:pote1}
and fall in one of the vacua (\ref{eq:vac2}). Then the domain walls
with the tension $\sigma_1$ or $\sigma_2$ are formed.

Domain walls must decay before the relevant BBN epoch. To this end, we
introduce a R-symmetry violating interaction that lifts the degeneracy of the vacua: 
$W = c' \la W \ra \sum_i\Phi_i^3/M_G^3$ with $\la W \ra = m_{3/2} M_G^2$,
leading to
$V_A =  c \,m_{3/2}^2 \sum_i \Phi_i^3/M_G + {\rm h.c.},$
 where $c$ and $c'$ are coupling constants. For simplicity
 we take $c$ both real and negative so that  true minimum is given by
$\phi_{{\rm true \,min}} = ( v_2,\dots, v_2)$.
The decay temperature is then expressed by
\bea
\frac{T_d}{100 {\rm MeV}}&\sim& \left(\lambda N \right)^{-\frac{3}{8}}
 \left(\frac{c}{0.1}\right)^{\frac{1}{4}}
\left(\frac{m_{3/2}}{1{\rm TeV}}\right)^{\frac{1}{2}}
\left(\frac{m_0}{1 {\rm TeV}}\right)^{\frac{3}{4}}.
\eea
When the bias becomes comparable to the energy of the domain walls, 
the topological defects are no longer stable and 
decay into $\Phi$-particles in the true vacuum.
Not to spoil the success of the BBN, however, $\Phi$-particles
should decay into standard model degrees of freedom. We assume
that $\Phi_i$ weakly couples to some heavy vector-like quarks, which
enable $\Phi$-particles to radiatively decay into the standard model
gauge bosons as soon as the domain walls decay.
Since the $R$-parity of $\Phi$ is even, the decay processes into
the lightest supersymmetric particles (LSPs), which may
overclose the universe, can be avoided if the mass of $\Phi$ is smaller
than two times the LSP mass.
Thus our model of domain walls can naturally
satisfy the constraints necessary to induce successful dilution.

As we saw in the above discussions, the domain wall can be
a viable candidate for late-time entropy production. What differs
from the other candidates is that the energy density falls off very
slowly: $\rho_{DW} \propto a^{-1}$.
Such a distinctive feature can lead to another important cosmological
consequence: the overclosure limit on the 
axion decay constant can be considerably relaxed. 
If we do not assume entropy production after axion begins
the coherent oscillation, the axion decay constant  is constrained as
$F_a \lesssim 10^{12}{\rm GeV}$ not to overclose the universe.
This upperlimit is relaxed  to $10^{15} {\rm \, GeV}$,  if
the late-time entropy production due to
the decay of nonrelativistic particles occurs~\cite{Kawasaki:1995vt}. 
In the following, 
we show that the axion decay constant is further
relaxed and can be as large 
the Planck scale if the domain-wall induced
entropy production occurs well below the QCD scale.

The axion starts to oscillate when the Hubble parameter becomes
comparable to the mass: $3H_{\rm osc}\simeq m_a(T_{\rm osc})$. 
The axion mass $m_a$ depends on the temperature $T$ as~\cite{KT}
\beq
m_a(T) \simeq \left\{
\bear{cc}
 \displaystyle{0.1 m_a (\Lambda_{\rm QCD}/T)^{3.7}} &{\rm for}~~\displaystyle{T\gtrsim \Lambda_{\rm QCD}/\pi },\\ 
m_a &{\rm for}~~\displaystyle{T\lesssim \Lambda_{\rm QCD}\pi },
\eear
\right.
\eeq
where $\Lambda_{\rm QCD} \simeq 0.2 {\rm GeV}$, and $m_a \simeq \Lambda_{\rm QCD}^2/F_a$ is the axion mass at the zero temperature.
Since the number density of the axion decreases as $\propto a^{-3}$,
the energy density of the axion when the domain walls decay is
\beq
\rho_a|_{T=T_d} = \frac{1}{2} m_a\,( m_a(T_{\rm osc}) F_a^2 \theta^2)
\frac{H_d^6}{H_{\rm osc}^6},
\eeq
where we have used $H\propto a^{-1/2}$ when the domain walls dominate
the universe. $\theta \sim O(1)$ denotes the initial amplitude of the axion field 
in the unit of $F_a$. The axion-to-entropy ratio is then given by
\beq
\frac{\rho_a}{s}= 1.3 \times 10^2 \frac{F_a^6 \,\theta^2\, T_d^9}{\xi(T_{\rm osc})^5 \Lambda_{\rm QCD}^8 M_G^6},
\eeq
where $\xi(T) \equiv m_a(T)/m_a \leq 1$.
This must be smaller than the present value of the ratio of the critical density
to the entropy.
That is, the axion decay constant should satisfy
\bea
F_a &\lesssim& 3.6 \times 10^{18} {\rm \, GeV}\, \xi(T_{\rm osc})^{\frac{5}{6}}
\theta^{-\frac{1}{3}} \non\\
&&\! \times
\left(\frac{\Omega_a h^2}{0.14}\right)^{\frac{1}{6}}
 \left(\frac{\Lambda_{\rm QCD}}{0.2{\rm\, GeV}}\right)^{\frac{4}{3}}
  \left(\frac{T_d}{10{\rm \, MeV}}\right)^{-\frac{3}{2}}.
\eea
Therefore the domain-wall induced
entropy production opens up the axion window to the Planck scale,
if $\xi(T_{\rm osc})$ is close to $1$. In other words, the axion is a good
candidate for dark matter if $F_a \sim M_G$. 
We need to check that $T_{\rm osc}$ is smaller
than $0.1{\rm \, GeV}$ so that $\xi(T_{\rm osc})$ is  $\sim 1$. 
To this end, the evolution of the cosmic temperature before the decay of
domain walls must be specified. If  the decay is approximated to be
an exponential decay with a constant decay rate, we obtain
\beq
T_{\rm osc} \simeq 0.05 {\rm GeV} \left(\frac{T_d}{10{\rm \, MeV}}\right)^{0.26}
\left(\frac{F_a}{M_G}\right)^{-0.13} \lesssim 0.1{\rm \, GeV}.
\eeq
Therefore, $\xi(T_{\rm osc})$ is close to $1$ when $T_d \sim 10{\rm \, MeV}$
and $F_a \sim M_G$.

In this letter we have investigated the dilution of unwanted 
cosmological relics
such as moduli by entropy production of the domain wall network.
It has been shown that the frustrated domain walls quickly
dominate the density of the universe and their decay produces
entropy large enough to dilute the dangerous moduli. 
We have also given a concrete model which leads to frustrated 
network of domain walls.
Moreover, we have found that the late-time decay of the domain 
walls greatly relaxes the constraint on the axion model and the axion 
decay constant $f_{a}$ as large as the Planck scale is allowed. 

Up to here,
we have assumed fully frustrated domain walls, 
however, this assumption can be
weaken to some extent. Similar arguments show that domain walls
whose energy evolves as $\rho_{DW} \propto a^{-1-\gamma}$ with
$\gamma \lesssim 0.2 (0.1)$ for $\kappa =0.2 (10^{-10})$ works as well.
Finally we make a comment on baryon number generation in the
present model. 
The domain-wall decay also dilute pre-existing baryon number. Therefore, there should be large baryon asymmetry before the
entropy production or baryon number should be generated after 
the decay of the domain wall. Since the decay temperature 
is expected to be low ($\sim 10$~MeV), it is unlikely to
produce the baryon number after the decay. Then the most
promising candidate for baryogenesis is the Affleck-Dine mechanism.
In order to produce sufficient baryon asymmetry  
the gravitino mass should be relatively small,  
$m_{3/2}\lesssim 10$~MeV~\cite{Asaka:1999xd},
as in the gauge-mediated SUSY breaking scenarios~\footnote{
For such small $m_{3/2}$, $\lambda$ ($m_0$) should be so small (large) that
the tension is larger than $(1 {\rm\, TeV})^3$. See Eq.~(\ref{eq:tension1}).}.
In the case of the thermal inflation, the Affleck-Dine mechanism
does not work due to $Q$-ball formation~\cite{Kasuya:2001tp}.
The crucial point is that the thermal inflation requires 
large messenger scale, which leads to large $Q$-balls and small
baryon number in the background universe. 
However, our model allows relatively small messenger scale,
since $v_2$ is much smaller than the vev of the flaton. Thus
the Affleck-Dine baryogenesis does work in the present scenario.

{\bf Acknowledgment:} 
 F.T. is grateful to K. Ichikawa and M. Kakizaki for useful discussions. 
 F.T.  would like to thank the Japan Society for
Promotion of Science for financial support.

\end{document}